\pdfoutput=1

\documentclass[openacc]{rstransa}

\newtheorem{theorem}{\bf Theorem}[section]

\usepackage{bm}

\titlehead{Research}

\usepackage[utf8]{inputenc}    
\usepackage[T1]{fontenc}       
\usepackage[english]{babel}    

\usepackage{tikz}
\usetikzlibrary {arrows.meta}
\numberwithin{figure}{section}
\numberwithin{equation}{section}
\newtheorem{definition}[theorem]{Definition}
\newtheorem{claim}[theorem]{Claim}
\usepackage{hyperref,bookmark,footnotebackref}
\hypersetup{hidelinks}
\hypersetup{bookmarksnumbered}

\newcommand \Tr {\operatorname{Tr}}
\newcommand \Scal 	{\mathcal S} 
\newcommand \Vcal 	{\mathcal V}
\newcommand \del	\partial
\newcommand \auth 	{\textsc}   
\newcommand \Acal 	{\mathcal A}   
\newcommand \Mcal 	{\mathcal M}   
\newcommand \Hcal 	{\mathcal H}   
\newcommand \Ecal 	{\mathcal E}   
\newcommand \RR 	{\mathbb R}   
   
\newcommand \eps 	\epsilon  
\newcommand \be 	{\begin{equation}}
\newcommand \ee 	{\end{equation}} 
\newcommand \bel 	{\be \label}
\newcommand \sgn 	{\operatorname{sgn}}

\newcommand \phase {{\boldsymbol \iota}}
\newcommand \phaseI {\mathbf{I}}
\newcommand \phaseII {\mathbf{II}}
\newcommand \tV {\widetilde{V}}

\newcommand \tmu {\widetilde{\mu}}
\DeclareMathOperator{\Span}{Span}

\newcommand{\Kcirc}{\mathring{K}}

\newcommand \rouge {} 

\begin{document}

\title{Scattering laws for interfaces in self-gravitating matter flows}

\author{Bruno Le Floch$^{1}$ and Philippe G. LeFloch$^2$}

\address{$^{1}$ Laboratoire de Physique Théorique et Hautes Énergies \& 
Centre National de la Recherche Scientifique, 
Sorbonne Université, 4 Place Jussieu, 75252 Paris, France.
{\small Email: {bruno@le-floch.fr}}
\\
$^{2}$ Laboratoire Jacques-Louis Lions \& 
Centre National de la Recherche Scientifique, 
Sorbonne Université, 4 Place Jussieu, 75252 Paris, France.
\\
{\small Email: {contact@philippelefloch.org}}}

\subject{general relativity, cosmology, fluid mechanics}


\corres{P.G. LeFloch\\
\email{contact@philippelefloch.org}
}

\emergencystretch=2em

\begin{fmtext}
We consider the evolution of self-gravitating matter fields that may undergo phase transitions, and we connect ideas from phase transition dynamics with concepts from bouncing cosmology. Our framework introduces \emph{scattering maps} prescribed on two classes of hypersurfaces: a gravitational singularity hypersurface and a fluid-discontinuity hypersurface. By analyzing the causal structures induced by the light cone and the acoustic cone, we formulate a local evolution problem for the Einstein--Euler system in the presence of such interfaces. We explain how suitable \emph{scattering relations} must supplement the field equations in order to ensure uniqueness and thus yield a complete macroscopic description of the evolution.

This viewpoint builds on a theory developed in collaboration with G.~Veneziano for quiescent (velocity-dominated) singularities in solutions of the Einstein equations coupled to a scalar field, where the passage across the singular hypersurface is encoded by a \emph{singularity scattering map}. The guiding question is to identify junction prescriptions that are compatible with the Einstein and Euler equations, in particular with the propagation of constraints. The outcome is a rigid set of \emph{universal} relations, together with a family of model-dependent parameters. Under physically motivated requirements (general covariance, causality, constraint compatibility, and ultra-locality), we aim to classify admissible scattering relations arising from microscopic physics and characterizing, at the macroscopic level, the dynamics of a fluid coupled to Einstein gravity.
\end{fmtext} 

\maketitle


\section{Introduction}
\label{section-1}

\paragraph{Purpose of this paper.}

We are interested in the modeling of interfaces propagating in a self-gravitating fluid, including shock waves, phase boundaries, or geometric singularities. We focus on \emph{sharp} interfaces represented by hypersurfaces across which the solutions to the Einstein--Euler equations admit distinct limits from both sides or, more generally, enjoy specified asymptotics. More broadly, we are interested in fluid-geometry interactions.
We explain here how to merge together selected ideas from bouncing cosmology and phase-transition dynamics. 

Many applications of the Einstein--Euler equations require continuum physics modeling \emph{beyond} the standard hydrodynamical equilibrium description and involve complex matter fields that possibly undergo phase transitions described by non-convex constitutive laws. The need to rely on \emph{physically realistic} laws ---beyond the standard polytropic perfect gases and including non-convex laws--- is well recognized within the astrophysics and numerical relativity communities; {\rouge 
see~\cite{Berbel-Serna,Berbel-Serna-Marquina,Font,FP,Kim,Kotake,Marquina-Serna-Ibanez,Rivieccio-Guerra-Ruiz-Font}.} While mathematical properties of shock waves in non-relativistic fluids were long studied in classical textbooks such as that of Courant and Friedrich~\cite{CourantFriedrichs}, the mathematical study of relativistic fluids and the properties of shock waves is comparatively more recent, going back to Lichnerowicz's pioneering work~\cite{Lichne}. On the other hand, many advances beyond the standard theory of shock waves were made in the recent two decades in order to encompass matter fields with complex dynamical behavior ---a topic which is still in very active development; {\rouge cf.~the monographs \cite{LeFloch-book,LeFlochThanh} together with \cite{AK1,AK2,BertozziShearer,FanSlemrod,FanSlemrod2,LeFloch-Oslo,LeFlochMohamadian,Shearer1986,Truskinovsky}. }

We tackle one of the challenging issues posed by the dynamics of complex matter fields, namely the \emph{formulation of junction conditions} at interfaces in self-gravitating matter. We distinguish two main regimes, depending on whether the spacetime in the neighborhood of the interface has \emph{bounded geometry} or \emph{singular geometry}, in a sense made precise below. Our objective is to propose a notion of \emph{scattering map} at such interfaces and to seek a \emph{classification} of such maps under physically motivated assumptions (locality, quiescence, causality, Lorentz invariance, etc.). While we develop here a theory of junction conditions describing the effect of gravity on relativistic fluids and scalar fields, the importance of junction relations was first recognized in non-relativistic fluid dynamics and in material science (overviewed in~\cite{LeFloch-book,LeFlochThanh}). As we argue below, without imposing junction relations, a sharp interface model would be \emph{incomplete}, and the initial value problem may fail to be well posed. 

{\rouge

Our key advance is to provide a unified setting encompassing shock waves, phase
interfaces, and spacelike singular hypersurfaces. In this sense, the
present paper may be viewed as placing several
distinct continuation problems within a common geometric formalism,
suitable both for mathematical analysis and for applications in
relativistic cosmology and astrophysics.

}

{\rouge

\paragraph{On non-convex equations of state and nonclassical waves.}

The present analysis is motivated by situations in
relativistic continuum physics in which the field equations cease, by
themselves, to determine how a solution should be continued across a
distinguished hypersurface. This issue is familiar in classical fluid
dynamics: once discontinuities form, weak solutions are not selected by
the conservation laws alone, and one must supplement the equations with
additional admissibility criteria. In the relativistic setting, an
analogous difficulty appears in a more geometric form. The interface may
represent a shock, a phase boundary, the onset of concentration of
matter, or a spacelike hypersurface across which the metric itself is
continued only in a singular sense. In all such cases, one is naturally
led to ask which data are transmitted across the interface and which
effective law governs the transmission.

In fluid dynamics (whether relativistic or not) with a convex equation of state, shock waves are fully constrained by the Rankine–Hugoniot conditions supplemented by Lax-type entropy criteria. The present paper does not assume any specific form of the equation of state and is therefore not restricted to the convex case. It is well known, however, that when convexity fails, as is expected in the vicinity of phase transitions in relativistic fluids, the structure of solutions to the Riemann problem changes substantially. In particular, non-convex equations of state may give rise to composite waves, undercompressive shocks, and phase boundaries that are not selected by the standard Rankine–Hugoniot and Lax conditions. In such situations, additional selection principles are required, often formulated in terms of kinetic relations or nucleation criteria, in order to restore uniqueness of weak solutions; cf.~ \cite{LeFloch-book,LeFlochThanh}. To this end, we formulate in this paper a general and covariant framework for such \emph{fluid scattering laws} within relativistic fluids. The explicit construction of scattering laws associated with specific non-convex equations of state, and their relation to concrete kinetic relations arising from microphysical models, is left for future work.
}

{\rouge

\paragraph{Effective description of bouncing cosmology scenarios.}

The main paradigm in describing early Universe cosmology is to have an initial inflationary period following an initial Big Bang singularity to account for large-scale homogeneity and isotropy.  Our work concerns a large family of alternative proposals whereby the Universe has a contracting phase---with an arbitrarily long time to homogenize---until reaching a highly compressed large curvature region, in which some unknown higher-curvature or quantum corrections to general relativity causes the Universe to expand, up to the present time.  This includes the string-inspired pre-Big-Bang scenario~\cite{Gasperini:1992em,Gasperini:2002bn,Gasperini:2023tus}, ekpyrotic universe~\cite{KOST}, slow contraction~\cite{Cook:2020oaj,Ijjas:2020dws,Kist:2022mew}, loop quantum cosmology~\cite{Ashtekar:2009a,Ashtekar:2009vc}, other modified gravity scenarios~\cite{De-Cesare}, matter bounces~\cite{Vitenti:2026hgs}, and more (see, e.g., the review~\cite{Brandenberger:2016vhg}).

These scenarios typically reduce to Einstein equations in both the contraction and expansion phases, coupled for instance to a fluid.
The intermediate stage is precisely where the classical Einstein equations are no longer expected to provide a complete description.
We adopt a more economical viewpoint, whereby the bounce is modeled by an interface carrying an effective law relating the incoming geometric and fluid variables to the outgoing ones, rather than resolving the microscopic physics of the high-curvature region. (This is similar to considering shock waves rather than their regularization by viscosity or capillarity.)

For this second class of interfaces, our formalism isolates the macroscopic consequences of microscopic effects that trigger the bounce. The requirement of \emph{ultra-locality} (that the scattering law involves only data at a single spacetime point on the interface) allows for a surprisingly small set of degrees of freedom. In fact, the bounce universally rescales the densitized anisotropy and acts on the scalar or fluid sector through a \emph{matter map}, with a possible dependence on a finite set of scalar invariants of the incoming extrinsic curvature---which is precisely where model dependence can enter---while the remaining outgoing state is fixed by consistency with the Einstein constraints.
These scattering laws are counterparts to those for fluid interfaces, and feature the same separation between universal laws analogous to Rankine--Hugoniot relations and model-dependent high-curvature effects.
They provide the missing information needed to relate incoming and outgoing states across the bounce hypersurface.
}

{\rouge

\paragraph{Relation with existing junction formalisms.}

Classical approaches to junction conditions in general relativity, such as the Darmois--Israel formalism and its extensions, describe the matching of spacetime regions across hypersurfaces, possibly supporting distributional sources. In this setting, the discontinuity of the second fundamental form is related to a surface stress–energy tensor, thereby providing a geometric characterization of thin shells and related configurations.

While these frameworks are well suited to the treatment of distributional solutions of Einstein’s equations, they are formulated primarily at the level of geometric compatibility conditions and do not, in general, provide a mechanism for selecting a unique evolution across the interface. They do not address situations in which additional physical input is required to relate incoming and outgoing states, as is the case for fluid discontinuities, phase transitions, or spacelike singular hypersurfaces.
In fact, we use these matching conditions in \autoref{section-5} to \emph{complement} the fluid scattering laws developed in \autoref{section-4}, in order to describe shock waves and phase transitions in the full Einstein--Euler system describing self-gravitating relativistic fluids.

Another limitation of the Darmois--Israel type junction conditions is that they concern interfaces where the metric remains \emph{bounded}. In the effective description of cosmological bounces, the metric is instead \emph{singular}, and the junction conditions must be formulated on suitable asymptotic expansions of geometric quantities.

}


\paragraph{Outline of this paper.}

{\rouge

To summarize, cosmological bounces and fluid interfaces
belong to the same general class of problems. In each case, the
evolution reaches a hypersurface across which the standard equations
must be complemented by additional information in order to restore
uniqueness. The point of view developed in the present paper is that this
additional information is most naturally encoded by a scattering law.
Such a law does not compete with the field equations; rather, it closes
them at the interface. It specifies how the relevant state variables are
matched across the hypersurface, while leaving the bulk evolution on
either side governed by the usual Einstein--Euler dynamics.

We focus here on the \emph{problem of defining and classifying} junction conditions for sharp interfaces, which is at once physically (and numerically) indispensable and mathematically tractable. The actual derivation of specific junction relations ---\emph{among} the allowable classes of junctions--- should be investigated next on a case-by-case basis for each application, and is beyond the scope of this paper. The specification of the junctions should be based on \emph{small-scale} physical modeling \emph{beyond} the Einstein--Euler model.

This paper is organized as follows. 
In \autoref{section-2}, we present the scattering laws for gravitational singularities with a \emph{scalar field}, understood in our previous works with G.~Veneziano. In \autoref{section-3}, we introduce a two-phase relativistic fluid model and next, in \autoref{section-4}, we define the notion of scattering maps for \emph{undercompressive} interfaces in relativistic two-phase fluids, generalizing the theory established for non-relativistic fluids.  We complement these with Darmois--Israel conditions to define scattering maps for such interfaces in self-gravitating fluids in \autoref{section-5}, and we then move on to scattering maps for geometric singularities, culminating in a classification result in Theorem~\ref{thm:fluid-scattering-map}. We close in \autoref{section-6} with observations toward a full classification of interface conditions and envision further applications.
}


\section{Scattering maps for Einstein--scalar geometric singularities}
\label{section-2}

\paragraph{Notion of singularity scattering map.}

The standpoint proposed in the present paper extends our earlier results in~\cite{LLV-1a,LLV-1b,LLV-2} on quiescent singularity hypersurfaces for self-gravitating, minimally coupled scalar fields. We first recall the relevant framework and statements, in a unified treatment of both spacelike and timelike singularity hypersurfaces. Specifically, we consider a Gaussian foliation $\bigcup_s \Hcal_s$ of a neighborhood of the singularity by hypersurfaces $\Hcal_s$, where $s$ denotes proper time in the spacelike case and proper distance in the timelike case. We write $g_{ab}=g_{ab}(s,\cdot)$ for the induced metric on $\Hcal_s$ (Riemannian if $\Hcal_s$ is spacelike and Lorentzian if $\Hcal_s$ is timelike), denote by $K_{ab}$ its second fundamental form and by $K_a^{\ b}=g^{bc}K_{ac}$ the associated mixed tensor, and write $\phi=\phi(s,\cdot)$ for the scalar field. Latin indices $a,b,\ldots$ range in $1,\dots,d$. The evolution and constraint equations in ADM variables are
\bel{adm-eqs}
\begin{aligned}
\del_s g_{ab} + 2K_{ab} &= 0 ,
& (\Tr K)^2-\Tr(K^2)-(\del_s\phi)^2&=\mp(R-\del_a\phi\,\del^a\phi),\mspace{-10mu}
\\
-\del_s^2\phi+\Tr(K)\,\del_s\phi &=\mp\nabla_a\nabla^a\phi,
& \nabla_aK^{a}_{\ b}-\del_b(\Tr K) + \del_s\phi\,\del_b\phi & = 0,
\\
\del_s K_a^{\ b}-(\Tr K)\,K_a^{\ b}&=\pm(R_a^{\ b}-\del_a\phi\,\del^b\phi),\mspace{-10mu}
\end{aligned}
\ee
with the upper sign in the spacelike case and the lower sign in the timelike case. Here, $R_a^{\ b}$ and $R$ are the Ricci and scalar curvature of $g$, and $\nabla$ is its Levi–Civita connection. We assume that $s=0$ labels the singularity hypersurface, denoted by $\Hcal_0$.

In the quiescent regime~\cite{Barrow:1978},
by definition, derivatives tangent to the leaves $\Hcal_s$ are negligible to leading order as $s\to 0$, and we may introduce the asymptotic profile\footnote{The profiles are intended to describe mesoscopic scales: $|s|$ should be small enough for the Einstein–scalar dynamics to enter the quiescent regime, but not so small that higher-curvature corrections or other microscopic physics become important. An arbitrary scale $s_*$ is fixed once and for all: changing $s_*$ redefines $(g^{\pm},K^{\pm},\phi_0^{\pm},\phi_1^{\pm})$ and scattering maps in an obvious way.} $(g_*^\pm(s),K_*^\pm(s),\phi_*^\pm(s))$ on the two sides $s\to 0^\pm$:
\be\label{gKphi-star}
g_*^{\pm}(s) = e^{2(\log|s/s_*|)\,K^{\pm}}\,g^{\pm},\qquad
K_*^{\pm}(s) = -\frac{1}{s} K^{\pm}, \qquad
\phi_*^{\pm}(s) = \phi_0^{\pm}\log|s/s_*|+\phi_1^{\pm},
\ee
where $g^\pm$ is a Riemannian (resp.\ Lorentzian) metric, $K^\pm$ is a $(1,1)$-tensor with trace $\Tr K^\pm=1$ and symmetric with respect to~$g^\pm$, and
$\phi_0^\pm,\phi_1^\pm$ are scalar functions on the limiting hypersurface. 

The asymptotic data must satisfy \emph{reduced constraints} (coming from the ADM constraints at leading order), namely
\bel{reduced-constraints}
1- K^{\pm}{}_b^{a}\,K^{\pm}{}_a^{b}=(\phi_0^\pm)^2,
\qquad
\nabla_a^{\pm}K^{\pm}{}_b^{a}=\phi_0^\pm\,\del_b\phi_1^\pm.
\ee
A \underline{scattering law} $(g^+,K^+,\phi_0^+,\phi_1^+) = \Scal(g^-,K^-,\phi_0^-,\phi_1^-)$ is defined through a scattering map~$\Scal$ that maps a singular data set to another satisfying the reduced constraints. Its physical content and interpretation depend on the signature: it corresponds to a \emph{bouncing law} across a spacelike \emph{crushing singularity} $\Hcal_0$, and to a \emph{transmission law} through a timelike \emph{defect (or wall)} when $\Hcal_0$ is timelike.


\paragraph{Spacelike singularities.}

Quiescent spacelike singularities are the classic asymptotically Kasner singularities.
The eigenvalues $k_1^\pm,\dots,k_d^\pm$ of~$K^\pm$, called Kasner exponents, describe how spatial directions scale with a power of the proper time~$s$.
They are expected to be dynamically stable under inhomogeneous perturbations provided the 
\emph{quiescence condition}
\be
\max_{i\neq j}(k_i+k_j)<1+\min_i k_i
\ee
holds. Many physical models (such as modified gravity, string-inspired models, loop quantum cosmology, etc.\@) have been proposed, for which the crushing singularity is replaced by a cosmological bounce described at mesoscopic scales by the asymptotic profiles~\eqref{gKphi-star} for ${s\lessgtr 0}$.
The physical model can be abstracted away into a scattering map~$\Scal$ that expresses the outgoing singular data in terms of the incoming data, thus allowing for the evolution problem to be well-posed if~$\Scal$ preserves the reduced constraints and quiescence condition. 


\paragraph{Timelike singularities.}

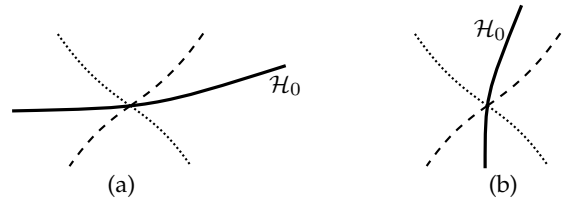
\begin{figure}
\centering
\begin{tikzpicture}[scale=0.8]
  \draw[very thick] (-1.8,-.05) .. controls (.5,.0) .. (2.7,.7) node [below] {$\Hcal_0$} coordinate [pos=.4] (A);
  \draw[thick, densely dotted] (A) to [bend left=10] +(-1.2,1.2);
  \draw[thick, densely dotted] (A) to [bend left=10] +(1.,-1.);
  \draw[thick, dashed] (A) to [bend right=10] +(1.2,1.2);
  \draw[thick, dashed] (A) to [bend right=10] +(-1.,-1.);
  \node at (0,-1.3) {(a)};
\end{tikzpicture}
\hspace{1.2cm}
\begin{tikzpicture}[scale=0.8]
  \draw[very thick] (0,-1.2) .. controls (0,.0) .. (.6,1.5) node [below left=2pt] {$\Hcal_0$} coordinate [pos=.4] (A);
  \draw[thick, densely dotted] (A) to [bend left=10] +(-1.2,1.2);
  \draw[thick, densely dotted] (A) to [bend left=10] +(1.,-1.);
  \draw[thick, dashed] (A) to [bend right=10] +(1.2,1.2);
  \draw[thick, dashed] (A) to [bend right=10] +(-1.,-1.);
  \node at (.3,-1.5) {(b)};
\end{tikzpicture}

\caption{Light cones (dashed/dotted lines) crossing a singularity hypersurface (full line) transversally:
(a) spacelike hypersurface, (b) timelike hypersurface.}
\label{fig:0}
\end{figure}

Timelike singularities have received comparatively less attention in the mathematical and physical literature.
They arise, for instance, in a detailed description of the collision of sufficiently energetic plane gravitational waves, beyond the cosmological bounce regime~\cite{LLV-1b}.
Such singularities typically (for positive Kasner exponents) correspond to a pinching phenomenon at a spatial point reached at finite proper distance, where the metric ceases to be smooth.
Since the metrics $g^\pm(s)$ are Lorentzian, one of the eigenvalues $k_1^\pm$ of $K^\pm$ is distinguished by having a timelike eigenvector, and the bound $k_1^\pm<1$ implied by the reduced constraints (unless $\phi_0^\pm=k_2^\pm=\dots=k_d^\pm=0$) ensures that the light cones, i.e.~the characteristics associated with the Einstein equations, cross the singularity hypersurface \emph{transversally}; see \autoref{fig:0}.
This places quiescent singularities in the class of undercompressive interfaces we discuss later.

In particular, since causal observers can reach the $s=0$ singularity in finite proper time, a junction condition relating the two sides of the singularity is required. In this setting, the scattering map $\Scal$ is naturally expected to be invertible; moreover, if the underlying microscopic model is \emph{parity-invariant}, $\Scal$ should coincide with its inverse. The dynamical stability of timelike quiescent singularities is more delicate. They should be stable provided the so-called \emph{velocity-dominated} structure persists for perturbations. At the very least, we expect the same quiescence condition as for spacelike singularities to be necessary, since it prevents a \emph{power-law divergence} of linearized metric perturbations as $s\to 0$.


\paragraph{Classification for Einstein--scalar gravitational singularities.}

Our analysis of singularity scattering maps in~\cite{LLV-1a,LLV-1b,LLV-2} assumes (i)~general covariance, (ii)~compatibility with the constraints,
and (iii)~ultra-locality, meaning that at each point of the hypersurface the outgoing data depend only on the incoming data at that point (and not on derivatives along the hypersurface). Under these modeling hypotheses we separate \emph{universal} junction laws from \emph{model-specific} laws.

We find it convenient to decompose
$K^\pm=\frac{1}{d}\delta+\Kcirc^\pm$ (with $\Tr \Kcirc^\pm=0$), together with 
\be
r^\pm = r(\phi_0^\pm) = \bigl(1-\tfrac{d}{d-1}(\phi_0^\pm)^2\bigr)^{1/2}
= \bigl(\tfrac{d}{d-1} \Tr(\Kcirc^\pm)^2\bigr)^{1/2}
\in [0,1] .
\ee 
We proved that the densitized traceless part of the extrinsic curvature satisfies the scaling relation 
\bel{universal-rescaling-law}
\bigl(|g|^{1/2}\,\Kcirc\bigr)^{\!+}= \gamma\,\bigl(|g|^{1/2}\,\Kcirc\bigr)^{\!-},
\qquad \gamma\in\RR,
\ee
which can be read as a \emph{universal rescaling law} of anisotropy across a spacetime bounce.
For the scalar field, the bounce acts on the pair consisting of the scalar field and the conjugate momentum
$\pi_\phi\sim |g|^{1/2}\del_t\phi$ by a \emph{canonical transformation}
\be
(\pi_\phi,\phi)_-\ \mapsto\ (\pi_\phi,\phi)_+ \quad \text{preserving}\quad d\pi_\phi\wedge d\phi.
\ee
Equivalently, after rewriting in terms of the singular data $(\phi_0,\phi_1)$, the map
$\Phi:(\phi_0^-,\phi_1^-)\mapsto(\phi_0^+,\phi_1^+)$ must satisfy the (signed) \emph{symplectic Jacobian identity}
\bel{symplectic-Jacobian}
\det\!\begin{pmatrix}
\del_{\phi_0^-}\!\bigl(\phi_0^+/r^+\bigr) & \del_{\phi_0^-}\phi_1^+\\
\del_{\phi_1^-}\!\bigl(\phi_0^+/r^+\bigr) & \del_{\phi_1^-}\phi_1^+
\end{pmatrix}
=\epsilon\,\del_{\phi_0^-}\!\Bigl(\frac{\phi_0^-}{r^-}\Bigr)
=\frac{\epsilon}{(r^-)^{3}},
\qquad \epsilon= \sgn(\gamma)\in\{0,\pm1\}.
\ee
The admissible scattering maps then split into two families.

\smallskip

\emph{(A) Anisotropic ultra-local scattering.}
There exists a (model-dependent) map~$\Phi$ satisfying~\eqref{symplectic-Jacobian}, and possibly depending on the scalar invariants\footnote{For timelike singularities, one of the exponents is singled out by having a timelike eigenvector.  This yields more invariants.}
$\chi_m = \Tr(\Kcirc^-/r^-)^{m}$ ($3\le m\le d$), such that
\be\label{Sani-expr}
(\phi_0^+,\phi_1^+)=\Phi(\chi_m,\phi_0^-,\phi_1^-),
\qquad
\Kcirc^+=\epsilon\,\frac{r^+}{r^-}\,\Kcirc^- .
\ee
Moreover, the metric after the bounce $g^+$ is obtained from $g^-$ by an \emph{anisotropic scaling} in the eigenframe of $\Kcirc^-$; more precisely,
$\log\bigl(g^+(g^-)^{-1}\bigr)$ is a polynomial in $\Kcirc^-$ of degree $<d$ whose coefficients are functions of $(\phi_0^\pm,\phi_1^\pm,\chi_m)$ explicited in terms of the parameter $\gamma\neq 0$ and of suitable derivatives of~$\Phi$.

\smallskip

\emph{(B) Isotropic ultra-local scattering.}
The choice $\gamma=0$ forces $\Kcirc^+=0$ and thus $K^+=\delta/d$. The reduced constraint then fixes
$\phi_0^+= \pm \sqrt{(d-1)/d}$, $\phi_1^+=\varphi\in\RR$,
while the outgoing metric can be any positive-definite rescaling of $g^-$ compatible with ultra-local covariance:
\be\label{Siso-expr}
\Scal_{\mathrm{iso}}:\ (g^-,K^-,\phi_0^-,\phi_1^-)\mapsto
\bigl(\Delta(\Kcirc^-,\phi_0^-,\phi_1^-)\,g^-,\ \delta/d, \epsilon \sqrt{(d-1)/d},\ \varphi\bigr),
\ee
where $\Delta=\sum_{n=0}^{d-1}\Delta_n(\phi_0^-,\phi_1^-,\chi_m)(\Kcirc^-)^n$ is required to have positive eigenvalues.
This family corresponds to strong information loss in anisotropy at the macroscopic level.


\paragraph{An explicit example.} 

As an illustration of our approach, it is insightful to consider a simple scenario in which an explicit scattering law can be obtained. For a \emph{pre-Big-Bang} cosmological scenario (a string theory-inspired bounce), the scalar field's momentum is mapped by the explicit Möbius-type rule:
\bel{preBigBang-map}
\phi_0^+ = - \frac{2\sqrt{d-1} + (d+1)\phi_0^-}{d+1 + 2d\,\phi_0^- / \sqrt{d-1}} .
\ee 
Together with the universal junction laws stated above, this prescription fixes the outgoing anisotropy amplitude and restricts the class of admissible post–bounce metrics~$g^+$.


\section{Hyperbolic model of a two-phase relativistic fluid}
\label{section-3}
 
\paragraph{Proposed model.} 

While our aim is to treat \emph{self-gravitating} fluids, we first introduce our model on a fixed background. We thus consider a globally hyperbolic Lorentzian manifold $(\Mcal, g^{(d+1)})$ of dimension $d+1$. We consider the initial value problem for the Euler equations on $(\Mcal, g^{(d+1)})$ 
\bel{equa-Euler}
\nabla^\alpha \big( T_{\alpha\beta}(\mu, u) \big) = 0, 
\ee
when the initial data for the density and the velocity are prescribed on a spacelike Cauchy hypersurface $\Hcal_0$. Here, $\mu \geq 0$ denotes the mass-energy density of the fluid and $u=(u^\alpha)$ its future-oriented timelike velocity, normalized by $u^\alpha u_\alpha = -1$. By convention, all Greek indices take the values $0, \ldots, d$. Whenever necessary indices are raised or lowered with the metric $g^{(d+1)} = g_{\alpha\beta} dx^\alpha dx^\beta$, and we use Einstein's summation convention. 

We consider a two-phase fluid, each phase being governed by the constitutive law of a perfect fluid. Specifically, in each phase $\phase = \phaseI, \phaseII$ the stress-energy tensor is 
\bel{eq:45}
T_{\phase, \alpha\beta} = \mu \, u_\alpha u_\beta + p_\phase(\mu) \, ( g_{\alpha\beta} + u_\alpha u_\beta ), 
\ee
where $p_\phase = p_\phase(\mu)$ is the pressure law in phase $\phase$. We assume that the associated sound speed 
\bel{equa-pressure} \textstyle
k_\phase := \sqrt{p_\phase'(\mu)} \in [0,1] 
\ee
is real and does not exceed the (normalized) light speed. We next introduce our model.

\begin{definition}
\label{def-the-models}
The \emph{two-phase relativistic model} represents a fluid flow on $\Mcal$ by a triple
$(\mu,u^\alpha,\phase)$, where $\mu:\Mcal\to[0,+\infty)$ is the mass-energy density,
$u=(u^\alpha)$ is a future-oriented unit timelike vector field, and the \emph{phase selector}
$\phase:\Mcal\to\{\phaseI,\phaseII\}$ takes only the two values $\phaseI$ or $\phaseII$.
In each phase $\phase=\phaseI,\phaseII$, the pressure is prescribed by a constitutive law
$p_\phase=p_\phase(\mu)$, and we define the global pressure field by
\be
p := p_{\phase}(\mu)\qquad \text{on }\Mcal.
\ee
The pair $(\mu,u^\alpha)$ is required to satisfy the relativistic Euler equations
\eqref{equa-Euler} with the stress-energy tensor \eqref{eq:45} associated with this pressure field.
\end{definition}
 
{\rouge

A two-phase relativistic fluid model provides an effective description of matter configurations in which two macroscopic phases, governed by distinct equations of state, are separated by a moving interface. Such configurations arise naturally in high-density astrophysical matter, for instance in idealized models of phase transitions inside compact stars.

}


\paragraph{Example 1. Strictly hyperbolic pressure laws.}

A standard class of constitutive assumptions is provided by the \emph{strict hyperbolicity conditions}
\bel{hyperbolic-eos}
\aligned
& 0 < p_\phase'(\mu) < 1 \quad (\text{for all } \mu>0), 
\qquad
p_\phase'(0)<1,
\endaligned
\ee
which ensure that the (relativistic) sound speed $k_\phase(\mu)=\sqrt{p_\phase'(\mu)}$ is real and strictly subluminal. For instance, an isothermal (affine) pressure law of the form
\be
p_\phase(\mu)=p_\phase^\star+k_\phase^2(\mu-\mu_\phase^\star),
\qquad \phase=\phaseI,\phaseII,
\ee
satisfies \eqref{hyperbolic-eos} for any constant sound speed $k_\phase\in(0,1)$.


\paragraph{Example 2. The stiff limit $k_\phase\equiv 1$.}

At the other end of the spectrum, we may consider the \emph{stiff} equation of state, for which the sound speed coincides with the (normalized) light speed:
\bel{hyperbolic-eos-1}
p_\phase(\mu)=p_\phase^\star+\mu-\mu_\phase^\star,
\qquad \phase=\phaseI,\phaseII.
\ee
In this case, the acoustic characteristics of the Euler system are tangent to the light cone, so that (linearized) fluid waves propagate at the same speed as gravitational waves. A key point, however, is that phase boundaries \emph{need not propagate at the characteristic speed} of the bulk phases: even when $k_\phase\equiv 1$, a phase interface selected by the Rankine--Hugoniot relations (and, in the undercompressive regime, by an additional kinetic/scattering relation) can move \emph{subsonically}, i.e.\ at a speed strictly smaller than~$1$. This phenomenon is central to the present discussion.
 

\paragraph{Example 3. Pressureless matter.}

Another class of particular interest in cosmology is the \emph{pressureless} regime
\bel{eq-920}
p_\phase(\mu)=0 \qquad \text{for all } \mu\ge 0.
\ee
In this case the flow behaves as \emph{dust}. In particular, the Euler system may admit
interfaces carrying \emph{surface layers}, in the sense that Dirac-type measure
contributions to the stress-energy tensor can propagate along hypersurfaces in
spacetime. This behavior can be ruled out by modifying the constitutive relation \eqref{eq-920}. For instance, Christodoulou~\cite{Christo-1992} points out that if the equation of state becomes \emph{stiff at
high density} ---that is, the sound speed reaches the light speed beyond a threshold--- 
then the formation of such surface layers can be suppressed.


\paragraph{Example 4. Discontinuous pressure laws.}

Equations of state that are \emph{discontinuous} functions of their argument are also relevant.
Specifically, we propose to consider a jump discontinuity at some threshold density $\mu_\star$,
\bel{equa-spacelike-jump}
p(\mu)
= \begin{cases}
p_\phaseI(\mu), \quad & \mu < \mu_\star,\\
p_\phaseII(\mu), \quad & \mu > \mu_\star,
\end{cases}
\ee
where $p_\phaseI$ and $p_\phaseII$ are monotone increasing functions on their respective domains.


\paragraph{Example 5. The low/high two-phase model.}

A fluid flow on $\Mcal$ is described by a pair $(\mu,u^\alpha)$, where the density is constrained to avoid a prescribed interval $(\mu_{\phaseI},\mu_{\phaseII})$, that is,
\[
\mu \in [0,\mu_{\phaseI}^*]\ \cup\ [\mu_{\phaseII}^*,+\infty).
\]
In this setup, the \emph{low-density phase}~$\phaseI$ is characterized by $\mu<\mu_{\phaseI}^*$, and the \emph{high-density phase}~$\phaseII$ by $\mu>\mu_{\phaseII}^*$. Accordingly, the phase selector $\phase$ in \autoref{def-the-models} is no longer an independent variable but is determined by the density, i.e.\ $\phase=\phase(\mu)$. The global pressure law then reads
\be
p(\mu)=
\begin{cases}
p_{\phaseI}(\mu), & \mu<\mu_{\phaseI}^*,\\
p_{\phaseII}(\mu), & \mu>\mu_{\phaseII}^*.
\end{cases}
\ee
Within the intermediate interval $\Ecal :=[\mu_\phaseI^*,\mu_\phaseII^*]$, the matter would presumably form a mixed (typically unstable) two-phase regime, in which the Euler system loses hyperbolicity and becomes of elliptic type; we therefore exclude this density range from the model. In particular, all admissible interfaces connect two states lying \emph{outside} the set $\Ecal$.

 
\paragraph{Klein--Gordon scalar fields.}

The same framework applies to minimally-coupled Klein--Gordon scalar fields, whose dynamics is governed by the Euler equations \eqref{equa-Euler} with stress-energy tensor
\bel{equa-Euler-phi}
\aligned
\nabla^\alpha \big( T_{\alpha\beta}(\phi) \big) 
 = 0, 
\qquad\quad 
T_{\alpha\beta}(\phi) 
 = \nabla_\alpha \phi \nabla_\beta \phi - \Big( \frac{1}{2}
g^{\gamma\delta} \nabla_\gamma \phi \nabla_\delta \phi + U(\phi) \Big) g_{\alpha\beta}. 
\endaligned
\ee
In particular, a massless scalar field can be reinterpreted as an \emph{irrotational stiff fluid} (i.e.\ with $p=\mu$ in the fluid variables), as recalled in \autoref{appendix-B}, below; it will be useful to keep this analogy in mind throughout the present discussion. Here, the potential $U=U(\phi)$ is a prescribed scalar function, for instance of quadratic type $U(\phi)\sim \phi^2$ or exponential type $U(\phi)\sim e^{\pm \phi}$.


\section{Scattering maps for interfaces in a two-phase relativistic fluid}
\label{section-4}

\paragraph{Rankine--Hugoniot junction conditions.}

We are primarily interested in physical interfaces separating two phases; however, our setup simultaneously encompasses interfaces within a single phase as a special case, and, in this case, such an interface is usually referred to as a shock wave. We thus assume that the flow variables $(\mu, u^\alpha)$ admit a jump discontinuity across a \emph{timelike hypersurface} $\Vcal_0$ in a spacetime $(\Mcal, g^{(d+1)})$.
For definiteness and without genuine loss of generality, we assume that one side belongs to the phase~$\phaseI$, while the other belongs to the phase~$\phaseII$\@.
The unit spacelike normal pointing towards the second side is denoted by~$V^\alpha$, and the relevant jump conditions along this hypersurface are derived by expressing the Euler equations \eqref{equa-Euler} in the distributional sense. It is a standard matter to derive the so-called Rankine--Hugoniot relations, by integration by parts of the Euler equations in a shrinking neighborhood of the interface.
Indeed, the fluid evolution equations \eqref{equa-Euler} provide us with relations between the fluid variables $(\mu_\phaseI, u_\phaseI)$ and $(\mu_\phaseII, u_\phaseII)$ evaluated on each side of the interface, namely
\be
V^\alpha \big( T_{\alpha\beta}(\mu_\phaseII, u_\phaseII) - T_{\alpha\beta}(\mu_\phaseI, u_\phaseI) \big) = 0.
\ee
Using the expression of the stress-energy tensor~\eqref{eq:45}, we arrive at the following definition.

\begin{definition} 
\label{def-junction-first}
For the two-phase models introduced in Definition~\ref{def-the-models}, the Rankine--Hugoniot \underline{junction conditions} at a fluid interface undergoing phase transitions, by definition, are 
\bel{equa-RHII}
\bigl( \mu_\phaseII + p_\phaseII(\mu_\phaseII) \bigr)  (u_\phaseII^\alpha V_\alpha) u_\phaseII^\beta 
- \bigl( \mu_\phaseI + p_\phaseI(\mu_\phaseI) \bigr)  (u_\phaseI^\alpha V_\alpha)  u_\phaseI^\beta 
+ \bigl(  p_\phaseII(\mu_\phaseII) - p_\phaseI(\mu_\phaseI) \bigr) \, V^\beta  = 0. 
\ee
\end{definition}

In \autoref{appendix-A}, we parametrize the solutions to~\eqref{equa-RHII}.
These conditions allow for \emph{contact interfaces} tangent to the fluid flow on both sides (namely $u_\phaseI\cdot V=u_\phaseII\cdot V=0$), and across which the parallel part of the fluid velocity is discontinuous, while pressure is continuous.  They also allow for \emph{shocks} or \emph{phase boundaries} with non-trivial jumps in all fluid variables.
We emphasize that, in \eqref{equa-RHII}, we allow the pressure laws to be \emph{different} on each side of the interface. As is well studied for non-relativistic fluids~\cite{AK1,AK2,BertozziShearer,FanSlemrod,FanSlemrod2,LeFloch-book,LeFloch-Oslo,LeFlochThanh,Shearer1986,Truskinovsky}, the dynamics of such interfaces is very different from that of single-phase interfaces. On the one hand, the space of solutions to the Rankine--Hugoniot relations may have 
a \emph{non-monotone} behavior; we do not investigate this issue in the present paper. On the other hand, the \emph{causality properties} of two-phase interfaces differ significantly from the single-phase case, and this is the feature we shall explain now.


\paragraph{Characteristic speeds.}

We recall that the Euler equations form a symmetrizable hyperbolic system (at least away from the vacuum state $\mu=0$).
Under the assumption~\eqref{equa-pressure} that the speed of sound $k_\phase = \sqrt{p_\phase'(\mu_\phase)}\in [0,1]$ is bounded by that of light, let us now consider the behavior of \emph{linearized waves} propagating on either side of a discontinuity hypersurface.
The existence of such waves with Fourier parameter $\xi\in\RR^{1+d}$ is equivalent to the vanishing of the characteristic polynomial of the system, explicitly
\bel{Delta-phase-xi}
\Delta_\phase(\xi) = (\xi\cdot u_\phase)^{d-1} \bigl( k_\phase^2 \xi\cdot\xi + (k_\phase^2 - 1) (\xi\cdot u_\phase)^2 \bigr) ,
\ee
in phases $\phase = \phaseI,\phaseII$.
Under our assumption $0\leq k_\phase\leq 1$, this polynomial is hyperbolic with respect to any timelike vector~$e_0$ in the sense that the polynomial $\kappa\mapsto\Delta_\phase(\zeta-\kappa e_0)$ has only real roots, for any $\zeta\in\RR^{1+d}$.  In other words, the line $\zeta+\RR e_0\subset\RR^{1+d}$ intersects the variety $\{\xi|\Delta_\phase(\xi)=0\}$ at $d+1$ points (accounting for multiplicity).

For comparison with the non-relativistic case, it is convenient to fix a reference future timelike vector~$e_0$, normalized to have $e_0\cdot e_0=-1$, and to decompose $V=V^0e_0 + V^1 e_1$ in terms of a unit spatial vector~$e_1$ orthogonal to~$e_0$, uniquely determined by the conditions $V^0=-e_0\cdot V$ and $V^1=|V-V^0e_0|>0$.  In this frame, waves propagating in a direction normal to the interface are described by $\xi\in\Span\{e_0,e_1\}$, which we normalize as $\xi=\lambda e_0+e_1$.
In the state $(\mu_\phase,u_\phase)$ on one side of the interface, the condition $\Delta_\phase(\xi)=0$ is satisfied by the following $d+1$ speeds~$\lambda$:
\bel{waves-multiplicity}
\aligned
\lambda_\phase^0 & = - u_{\phase 1}/u_{\phase 0} \quad \text{(multiplicity $d-1$)}, \\
\lambda_\phase^{L,R} & = - \frac{(1 - k_\phase^2) u_{\phase 0} u_{\phase 1} \pm k_\phase \sqrt{1 + (1 - k_\phase^2) |u_{\phase\parallel}|^2}}{k_\phase^2 + (1 - k_\phase^2) u_{\phase 0}^2} ,
\endaligned
\ee
where we decomposed the fluid velocity as $u_\phase = -u_{\phase 0}e_0+u_{\phase 1}e_1+u_{\phase\parallel}$ with $u_{\phase\parallel}$ orthogonal to $e_0,e_1$ (hence parallel to the interface), which obey the normalization $u_{\phase 0}^2 = 1 + u_{\phase 1}^2 + |u_{\phase\parallel}|^2$.
Besides the $d-1$ waves propagating at the same speed~$\lambda_\phase^0$ as the fluid, we can distinguish two families of waves, conveniently referred to as the left-hand and right-hand waves, with speeds $\lambda_\phase^L, \lambda_\phase^R$ corresponding to the choice of a plus sign and minus sign in the factor $\pm k_\phase$, respectively.
Under the strict hyperbolicity condition~\eqref{hyperbolic-eos}, namely $0<k_\phase(\mu)<1$ away from vacuum, the speeds differ from each other (and are real), so that the system is strictly hyperbolic.


\paragraph{Geometry of the characteristics.} 

Interfaces are distinguished according to how their velocity $z = V^0 / V^1$ compares with the wave speeds on the two sides $\phase=\phaseI,\phaseII$ for each family ($\lambda_\phase^L, \lambda_\phase^0, \lambda_\phase^R$).
We single out the most important types for our purposes ---while omitting certain cases of \emph{equalities} in some of the inequalities below which may occur at particular points on the Hugoniot curves. Moreover, some of these inequalities are included for clarity, despite being implied by the others since $\lambda_\phase^L<\lambda_\phase^0<\lambda_\phase^R$. In addition, interfaces having the \emph{opposite} strict inequalities are unphysical, and evolve as \emph{rarefaction waves} (not analyzed in the present text). 

\begin{itemize} 

\item {\bf Left-hand compressive interfaces}  are defined by
\be
\lambda^L_\phaseI > z > \lambda^L_\phaseII; 
\qquad
z < \min(\lambda^0_\phaseI, \lambda^0_\phaseII).
\ee 

\item {\bf Left-hand undercompressive interfaces}  are defined by
\be
\aligned
z < \min(\lambda^L_\phaseI, \lambda^L_\phaseII) 
\text{ or } 
z > \max(\lambda^L_\phaseI, \lambda^L_\phaseII);
\qquad
z < \min(\lambda^0_\phaseI, \lambda^0_\phaseII).
\endaligned
\ee

\item {\bf Contact interfaces}  are defined by
\be
z = - u_{\phaseI 1}/u_{\phaseI 0} = - u_{\phaseII 1}/u_{\phaseII 0} . 
\ee 

\item {\bf Right-hand compressive interfaces}  are defined by
\bel{right-hand-compressive-interface}
z > \max(\lambda^0_\phaseI, \lambda^0_\phaseII); 
\qquad 
\lambda^R_\phaseI > z > \lambda^R_\phaseII.
\ee 

\item {\bf Right-hand undercompressive interfaces}  are defined by
\be
\aligned
z > \max(\lambda^0_\phaseI, \lambda^0_\phaseII);
\qquad 
z < \min(\lambda^R_\phaseI, \lambda^R_\phaseII)
\text{ or }
z > \max(\lambda^R_\phaseI, \lambda^R_\phaseII).
\endaligned
\ee

\end{itemize}

\begin{figure}
  \centering
  \begin{tikzpicture}[scale=1.3]
    \draw[->](-1.3,0) -- (-1.3,.6) node [above] {$e_0$};
    \draw[->](-1.3,0) -- (-.7,0) node [above] {$e_1$};
    \draw[very thick] (-.2,-.2) .. controls (.75,.8) .. (1.2,2.0) coordinate [pos=.4] (A);
    \draw[thick, dashed] (A) to [bend right=10] (-.4,0) (A) to [bend left=10] (.2,0);
    \draw[thick, densely dotted] (1.3,0) to [bend right=10] (A) to [bend right=10] (-.1,1.4);
    \node at (.5,1.9) {$(\mu_\phaseI, u_\phaseI)$};
    \node at (1.75,1.9) {$(\mu_\phaseII, u_\phaseII)$};
    \node at (-.1,1.5) {$n_\phaseI=1$};
    \node at (.76,0) {$n_\phaseII=2$};
  \end{tikzpicture}\quad
  \begin{tikzpicture}[scale=1.3]
    \draw[very thick] (-.2,-.2) .. controls (.75,.8) .. (1.2,2.0) coordinate [pos=.4] (A);
    \draw[thick, dashed] (A) to [bend right=10] (-.4,0) (A) to [bend right=10] (1.35,1.4);
    \draw[thick, densely dotted] (1.3,0) to [bend right=10] (A) to [bend right=10] (-.1,1.4);
    \node at (.5,1.9) {$(\mu_\phaseI, u_\phaseI)$};
    \node at (1.75,1.9) {$(\mu_\phaseII, u_\phaseII)$};
    \node at (-.1,1.5) {$n_\phaseI=1$};
    \node at (.76,0) {$n_\phaseII=1$};
  \end{tikzpicture}\qquad
  \begin{tikzpicture}[scale=1.3]
    \draw[very thick] (-.2,-.2) .. controls (.75,.8) .. (1.2,2.0) coordinate [pos=.4] (A);
    \draw[thick, dashed] (A) to [bend left=10] (.2,0) (A) to [bend right=10] (.8,1.4);
    \draw[thick, densely dotted] (1.3,0) to [bend right=10] (A) to [bend right=10] (-.1,1.4);
    \node at (.5,1.9) {$(\mu_\phaseI, u_\phaseI)$};
    \node at (1.75,1.9) {$(\mu_\phaseII, u_\phaseII)$};
    \node at (.4,1.5) {$n_\phaseI=2$};
    \node at (.76,0) {$n_\phaseII=2$};
  \end{tikzpicture}
  \caption{Right-hand compressive (left figure) versus undercompressive (middle and right figures) interfaces. The dashed (resp.\ dotted) lines represent characteristic curves of the right-hand (resp.\ left-hand) family, and the thick line denotes the interface with speed~$z$.  Characteristics with speed $\lambda_\phase^0$ are omitted to avoid clutter.}
  \label{fig:2} 
\end{figure}
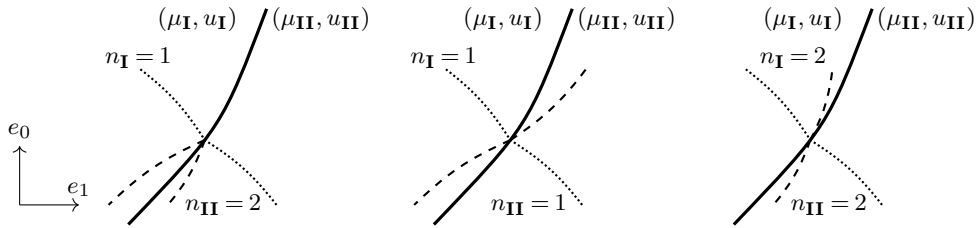

\noindent The third class of interfaces is linear in nature, and from now on we focus our attention on the properties of (compressive, undercompressive) nonlinear waves. The inequalities in the compressive interfaces are nothing but the usual Lax shock inequalities. We refer to \autoref{fig:2} for an illustration of right-hand compressive and undercompressive interfaces.
The formalism extends formally to spacelike interfaces such as the geometric singularities in \autoref{section-2}, by considering interface speeds $|z|>1$: such interfaces are undercompressive since all wave speeds lie on the same side of~$z$.

The compressive or undercompressive nature of an interface is \emph{independent} of our choice of reference future timelike vector~$e_0$.
Indeed, let us rephrase it in terms of the number $n_\phase(V)$ of solutions of $\Delta_\phase(V-\kappa e_0)=0$ with $\kappa\in[0,+\infty)$, accounting for multiplicities ---varying~$e_0$ among future timelike vectors does not change this number since the variety $\{\xi|\Delta_\phase(\xi)\}$ is spacelike or null.
The wave speeds~$\lambda$ are roots of $\Delta_\phase(\lambda e_0+e_1)$ and equivalently (by rescaling) of the polynomial $\Delta_\phase(V + V^1(\lambda-z)e_0)$, so that $n_\phase(V)$ counts the number of wave speeds less than the interface speed~$z$.
In this language, left-hand compressive interfaces have $n_\phaseI(V) = 0$ and $n_\phaseII(V)=1$, left-hand undercompressive interfaces have $n_\phaseI(V) = n_\phaseII(V) \in \{0,1\}$, while right-hand compressive interfaces have $n_\phaseI(V)=d$ and $n_\phaseII(V) = d + 1$ and right-hand undercompressive interfaces have $n_\phaseI(V) = n_\phaseII(V) \in \{d, d + 1\}$.
For contact interfaces, the counts $n_\phase$ degenerate as $V$~itself is a root of~$\Delta_\phase$.

A crucial difference between compressive and undercompressive interfaces lies in the \emph{number of outgoing degrees of freedom}, which must be determined from incoming data by suitable junction conditions.  This can be seen by counting characteristics.
As depicted in \autoref{fig:2}, the side~$\phaseI$ of the interface has $d+1-n_\phaseI(V)$ incoming and $n_\phaseI(V)$ outgoing characteristics, and the other side has $n_\phaseII(V)$ incoming and $d+1-n_\phaseII(V)$ outgoing ones, for a total of $d+1+n_\phaseI(V)-n_\phaseII(V)$ outgoing degrees of freedom.
In an evolution problem, the intersection of the interface with a time slice is known, but the interface speed normal to that locus is an additional scalar unknown.
These outgoing data are largely fixed in terms of incoming data by the $d+1$ Rankine--Hugoniot relations, leaving precisely $1+n_\phaseI(V)-n_\phaseII(V)$ \emph{undetermined degrees of freedom:} none in the compressive case and exactly \emph{one scalar unknown} for undercompressive interfaces.


\paragraph{Scattering maps for fluid phase interfaces.}

In the \emph{non-relativistic} literature on fluid dynamics and materials, \emph{compressive} discontinuities are selected dynamically by the inviscid Euler flow and may form from smooth initial data, whereas \emph{undercompressive} shocks or phase interfaces typically arise only as \emph{limits of augmented models} (for instance incorporating viscosity and capillarity) when small-scale parameters vanish. Both types of waves can be physically relevant and stable, but they are governed by different selection principles and therefore call for different analytical tools.

Motivated by junction laws proposed for liquid–vapor mixtures and austenite/martensite phase transitions~\cite{AK1,LeFloch-book,Truskinovsky}, we therefore supplement the relativistic Euler equations and Rankine--Hugoniot relations with an additional relation to obtain a \emph{scattering law} (classically a \emph{kinetic relation}) that determines the dynamics of an undercompressive phase boundary. This additional relation is intended to encode interfacial mechanisms—such as surface-tension-type effects and entropy production—that lie \emph{beyond} the inviscid Euler description.


We consider the set $\Acal$ of \emph{admissible tuples} $(\mu_\phaseI,u_\phaseI,\mu_\phaseII,u_\phaseII,V)$, characterized by
\be
\Acal :
\left\{
\aligned
(1)\ & \text{positivity } \mu_\phaseI, \mu_\phaseII \ge 0;
\\[-.5ex]
(2)\ & u_\phaseI, u_\phaseII \ \text{unit timelike and future-oriented};
\\[-.5ex]
(3)\ & V \ \text{unit spacelike};
\\[-.5ex]
(4)\ & \text{the Rankine--Hugoniot relations \eqref{equa-RHII} hold};
\\[-.5ex]
(5)\ & \text{the associated interface is undercompressive}.
\endaligned
\right.
\ee
These conditions are purely algebraic, in contrast with the differential Einstein constraints in Sections~\ref{section-2} and~\ref{section-5}. We provide a convenient parametrization of this set in~\autoref{appendix-A}.

Our goal is to fix the indeterminacy of undercompressive interfaces by introducing a suitable codimension~$1$ locus in~$\Acal$ expressed for some scalar function $\Pi:\Acal\to\RR$ as
\bel{locus-in-Acal}
\Pi(\mu_\phaseI, u_\phaseI, \mu_\phaseII, u_\phaseII, V) = 0 .
\ee
\emph{Lorentz invariance} severely restricts its functional form.  The law can only depend on densities and scalar invariants of the unit vectors $u_\phaseI,u_\phaseII,V$, namely $u_\phaseI\cdot u_\phaseII$, $V\cdot u_\phaseI$, $V\cdot u_\phaseII$.
These in turn are fixed by the Rankine--Hugoniot relations as explicit expressions 
(computed explicitly in~\eqref{dot-by-RH})
 in terms of the densities (and pressure laws) on both sides.
Thus, without loss of generality, the locus in~\eqref{locus-in-Acal} can be written as $\Psi(\mu_\phaseI, \mu_\phaseII)=0$ for some scalar function~$\Psi$.
Furthermore, this can generically be converted (locally in the space of states) to a bijective map expressing the fluid density on one side (say, phase~$\phaseII$) in terms of the other.
Alternatively, one could characterize this law by specifying the interface velocity as a function of the state $(\mu_\phaseI,u_\phaseI)$.

\begin{definition}
\label{def-oKDF9}
A \underline{fluid scattering law} relates the left- and right-hand states across an undercompressive fluid interface.
It is defined from the relation
\be
\mu_\phaseII = \varphi(\mu_\phaseI),
\ee
based on a prescribed map $\varphi:[0,+\infty)\to[0,+\infty)$, together with the Rankine--Hugoniot relations fixing all remaining degrees of freedom.
Given a $(d-1)$-dimensional time slice~$\Sigma$ of an interface and a state $(\mu_\phaseI,u_\phaseI)$ on one side, the density $\mu_\phaseII=\varphi(\mu_\phaseI)$ is determined, as well as the normal traces $V\cdot u_\phaseI$ and $V\cdot u_\phaseII$ made explicit in~\eqref{dot-by-RH}.  The value $V\cdot u_\phaseI$ uniquely determines the unit normal one-form~$V$ among unit vectors orthogonal to~$\Sigma$.  Finally, the Rankine--Hugoniot relation ${(\mu_\phaseII + p_\phaseII)  (V\cdot u_\phaseII) u_\phaseII} = {(\mu_\phaseI + p_\phaseI)  (V\cdot u_\phaseI)  u_\phaseI} - {(p_\phaseII - p_\phaseI) \, V}$ fixes the velocity.
\end{definition}

{\rouge

\paragraph{A relativistic analogue of kinetic relations.} 

For fluids with a non-convex equation of state, undercompressive interfaces can occur, and the classical Rankine--Hugoniot conditions generally do not select a unique weak solution. As in the theory of nonclassical shocks and phase boundaries (cf.~\cite{LeFloch-book,LeFloch-Oslo}), one must therefore add a selection principle, such as a kinetic relation and, possibly, a nucleation criterion. In this sense, the scattering law of Definition~4.2 is its relativistic analogue: it provides, in covariant form, the transmission rule that determines the outgoing state from the incoming one at the interface. This viewpoint also suggests extending to relativistic fluids the analytical and numerical constructions developed for non-relativistic kinetic relations (cf.~\cite{LeFloch-book,LeFlochThanh,LeFlochMohamadian}).

}


\section{Scattering maps for Einstein--Euler geometric singularities}
\label{section-5}
 
\paragraph{Bounded geometry.} 

{\rouge
We are now in a position to consider interfaces in self-gravitating fluids, across which the geometric and fluid variables may either blow up or remain bounded.
Junction conditions across these two types of interfaces play the same role of completing the field equations.  For geometric singularities, we explain momentarily how to extend our theory in~\cite{LLV-1a,LLV-1b,LLV-2} concerning the Einstein--scalar field system.
On the other hand, taking into account the effect of self-gravitation on the fluid interfaces described in \autoref{section-4} is direct. To ensure well-posedness of the initial value problem, it is necessary to specify all remaining degrees of freedom. 
As long as the geometry remains bounded, the Darmois--Israel junction conditions apply~\cite{Darmois,Israel}.
}

\begin{definition}\label{def-scattering-law}
In a self-gravitating fluid flow, the \underline{scattering law} across a general fluid interface (shock wave, phase boundary) consists by definition of the standard Rankine--Hugoniot conditions~\eqref{equa-RHII} and the Darmois--Israel conditions requiring continuity of the metric and the second fundamental form, as well as, in the undercompressive case, a fluid scattering law (Definition~\ref{def-oKDF9}).
\end{definition}

{\rouge

\paragraph{On Dirac mass concentrations and thin shells.}

An important feature of \emph{pressureless} relativistic fluid models~\eqref{eq-920}, already present in the non-relativistic theory, is the possibility of concentration effects in which part of the mass accumulates on hypersurfaces. In the distributional limit, this leads to Dirac measure components in the fluid density (and a discontinuous fluid velocity), corresponding to the appearance of \emph{thin matter shells}. Such configurations arise naturally in the evolution of pressureless fluids as well as through shock-wave interactions; they are consistent with a weak formulation of the Einstein–Euler system.
As a consequence of this surface stress tensor, the Darmois--Israel junction conditions impose that the spacetime metric remains continuous, while the second fundamental form contains a jump discontinuity.

Independently, there exist \emph{impulsive gravitational waves} across which the curvature also contains Dirac mass terms \cite{Penrose,MarsSenovilla}. (Cf.~also~\cite{LeFlochStewart-2011} for a mathematical study in plane symmetry). They represent geometric discontinuities propagating along null hypersurfaces. In both situations, the jump is governed by standard junction conditions and there is \emph{no need} to introduce additional scattering relations which apply to undercompressive phase boundaries.
From a computational perspective, the presence of Dirac mass concentrations poses significant challenges, since standard numerical schemes may fail to properly capture measure-valued components without appropriate regularization or interface-tracking techniques.

}

 
\paragraph{Quiescent singularities in an asymptotically stiff self-gravitating fluid.}

We now turn to spacelike\footnote{Timelike singularities are not governed by the same class of asymptotic profiles as~\eqref{gKmuu-star} below, as the fluid velocity~$u$ would then be spacelike near the singularity, instead of timelike as it must.} hypersurfaces where the geometric (and fluid) variables are singular.
Motivated by our treatment of fluids in previous sections, we emphasize that such interfaces are \emph{undercompressive}, in the sense that the characteristic directions cross the wave cone; cf.~\autoref{fig:0} (left) for an illustration.
In accordance to this, the scattering maps below are not entirely fixed by universal laws similar to the Rankine--Hugoniot relations, but retain a free functional degree of freedom~$\psi$, analogous to the map~$\varphi$ for phase boundaries in Definition~\ref{def-oKDF9}.

We assume that the fluid pressure $p=p(\mu)$ is \emph{asymptotically stiff}, in the sense that for some positive exponent $\eps>0$
\bel{equa-jdjh44}
p(\mu) = \mu + O(\mu^{1-\eps}) , \qquad \mu \to +\infty .
\ee
Hence, in the irrotational case the fluid behaves at high density like a scalar field; cf.~\autoref{appendix-B}.
Solutions with spacelike quiescent singularities for exactly stiff fluids were constructed by Andersson and Rendall~\cite{AnderssonRendall}. (See also the textbook~\cite{Rendall-book}.)
Their Fuchsian analysis readily applies to asymptotically stiff fluids, and the solutions $(g, K, \mu, u)$ of interest to us are close to asymptotic profiles for $t\gtrless 0$: 
\bel{gKmuu-star}
\aligned
g_*^{\pm}(t) & = e^{2(\log|t/t_*|)\,K^{\pm}}\,g^{\pm},\quad
& K_*^{\pm}(t) & = -t^{-1} K^{\pm},
\\
\mu_*(t) & = t^{-2} \mu^{\pm} , \quad
& u_{*a}(t) & = \frac{1}{2} t\log|t/t_*| \del_a (\log\mu^{\pm}) + t v^{\pm}_a , \quad a=1,\dots,d ,
\endaligned
\ee
and $u_{*0}(t)=-1$, with corrections in powers of~$t$.  Here the singular data sets $(g^{\pm},K^{\pm},\mu^{\pm},v^{\pm})$ consist of a Riemannian metric~$g^\pm$, a unit-trace tensor~$K^\pm$ that is symmetric with respect to~$g^\pm$, a non-negative scalar field $\mu^\pm\geq 0$, and a vector field on the singularity hypersurface~$\Hcal_0$.
They are subject to the reduced Hamiltonian and momentum constraints,
\bel{reduced-constraints-fluid}
1- K^{\pm}{}_b^{a}\,K^{\pm}{}_a^{b} = 2\mu^\pm ,
\qquad
\nabla_a^{\pm}K^{\pm}{}_b^{a} = 2\mu^\pm v^\pm_b .
\ee
The main difference compared to the scalar field case is that the source vector $2\mu^\pm v^\pm_b$ is no longer (a multiple of) a gradient.


\paragraph{Scattering maps for quiescent singularities.}

The outgoing singular data is expressed in terms of incoming data through a \underline{scattering map} $\Scal:\ (g^-,K^-,\mu^-,v^-)\longmapsto (g^+,K^+,\mu^+,v^+)$ that must respect the constraints.  We concentrate on maps~$\Scal$ that are \emph{ultra-local}, in the sense that they do not involve spatial derivatives along the hypersurface: this is relevant for the description of bounces caused by microscopic physics that respects the decoupling of spatial points in the quiescent regime.
In this context, we extend the analysis performed in~\cite{LLV-1a} for spacelike singularities and scalar fields. In this short presentation, we only outline the proof since it is analogous to the one therein ---although the set of tensor structures allowed by the presence of the vector field~$v^-$ alongside the $(1,1)$ tensor~$K^-$ complicates the analysis.

{\rouge

Lorentz covariance implies that the $(1,1)$ tensor $K^+$ is a linear combination
\bel{Kp-lin-combin}
K^+ = \sum_{m=0}^{d-1} a_m (K^-)^m + \sum_{l,m=0}^{d-1} a_{l,m} ((K^-)^lv^-)\otimes((K^-)^m v^-)^\#,
\ee 
where $\#$ denotes raising indices with the metric~$g^-$, and
with coefficients $a_m,a_{l,m}$ that depend on scalar invariants of $(g^-,K^-,\mu^-,v^-)$.  By positivity of the metrics, we can also write $g^+ = \exp(P(K^-,v^-)) g^-$, where the $(1,1)$ tensor $P$ decomposes as a linear combination of the tensor structures in~\eqref{Kp-lin-combin}, and $\exp$ denotes the matrix exponential.
By expressing the tensor $\nabla^+-\nabla^-$ (Christoffel symbols) in terms of $\nabla^- g^+$ and taking into account that $K^{+a}{}_b (g^+)^{-1\,bc}$ is symmetric in $a,c$, we then evaluate
\bel{diveKp-calc}
\nabla^+_a K^{+a}{}_b = \nabla^-_a K^{+a}{}_b - \frac{1}{2} \nabla^-_a(P^c{}_c) K^{+a}{}_b - \frac{1}{2} \omega K^{+a}{}_c \nabla^-_b P^c{}_a ,
\ee
which can then be further expanded by writing $K^+$ and~$P$ as linear combinations~\eqref{Kp-lin-combin}.
The scattering map should be such that this divergence is equal to the non-derivative term $2\mu^+ v^+_b$ whenever the (reduced) constraints~\eqref{reduced-constraints-fluid} are satisfied by the incoming data.  In particular, all derivatives of $K^-$ must \emph{cancel out} when evaluating~\eqref{diveKp-calc}, apart from the divergence $\nabla^-_a K^{-a}{}_b$ which is given by the momentum constraint.
Given all the independent tensor structures, this necessary cancellation can only occur if almost all coefficients $a_m,a_{l,m}$ vanish. This leads to the same universal law~\eqref{universal-rescaling-law} as for the Einstein--scalar system: 
\bel{universal-rescaling-law-repeat}
|g^+|^{1/2} \Kcirc^+ = \gamma |g^-|^{1/2} \Kcirc^- , \qquad \gamma\in\RR ,
\ee
where we recall that $\Kcirc^{\pm}=K^{\pm}-\delta/d$ are the traceless parts.
Since the case $\gamma=0$ is elementary, let us pursue the case $\gamma\neq 0$.
The symmetry of $g^+$ and of $\Kcirc^+g^+$ (hence of $\Kcirc^-g^+$) then forbids the tensor structures $((K^-)^lv^-)\otimes((v^-)^\#(K^-)^m)$ from appearing in~$g^+$, allowing only powers of~$K^-$.  Positivity of~$g^\pm$ then allows us to write $g^+=\exp(P(\Kcirc^-))g^-$ in terms of the matrix exponential~$\exp$ of some polynomial~$P$.
At this stage the divergence $\nabla^+_a K^{+a}{}_b$ reduces to a linear combination of terms of the form $(\text{scalar})\partial_b(\text{scalar})$ which must cancel out.  This fixes all the outgoing data in terms of only two scalar functions $\omega,\psi$ of the scalar invariants (note that $\tau_1^-=0$)
\bel{scalar-invariants-tau-ups}
\tau_m^-=\Tr\bigl((\Kcirc^-)^m\bigr) , \qquad
\upsilon_m^-=g^-\bigl((\Kcirc^-)^{m-1}v^-,v^-\bigr) , \qquad
m=1,\dots,d .
\ee
Requiring~$\Scal$ to be \emph{differentiable} at $\Kcirc^-=0$ imposes $\psi=0$ identically.
We reach a classification of such scattering maps.

}

\begin{theorem}\label{thm:fluid-scattering-map}
The (ultra-local) scattering maps for spacelike singularity hypersurfaces in a self-gravitating asymptotically stiff, compressible fluid are isotropic or anisotropic maps, defined as follows.

(A) \emph{Isotropic scattering maps} have vanishing $\Kcirc^+$ and~$v^+$, constant $\mu^+=(d-1)/(2d)$, and a metric~$g^+$ given by any positive Lorentz-covariant combination of $(g^-,K^-,\mu^-,v^-)$.

(B) \emph{Anisotropic maps} depend on a non-zero constant $\gamma\in\RR$ and a positive scalar function~$\omega = {\omega(\tau_2^-,\dots,\tau_d^-,\allowbreak\upsilon_1^-,\dots,\upsilon_d^-)}$ of the invariants~\eqref{scalar-invariants-tau-ups}.
The extrinsic curvature, metric, fluid momentum, and shifted fluid density undergo a conformal rescaling by a scalar factor: 
\bel{anisotropic-map-fluid}
\aligned
\Kcirc^+ = \gamma \omega^{-1} \Kcirc^- , \qquad
g^+ & = \omega^{2/d} g^- , \qquad
\mu^+ v^+ = \gamma \omega^{-1} \mu^- v^- ,
\\
\frac{d-1}{2d} - \mu^+ & = \gamma^2 \omega^{-2} \Bigl(\frac{d-1}{2d} - \mu^-\Bigr) .
\endaligned
\ee
\end{theorem}

{\rouge

\paragraph{Remark on the role of the fluid velocity.}

Compared with the scalar-field or vacuum case, the presence of a distinguished spatial velocity field~$v^{\pm}$ a priori allows additional tensorial structures to enter Lorentz-covariant ultralocal expressions that may serve as candidate scattering maps.
However, the preservation of momentum constraints---which features specifically the divergence of the (asymptotic) extrinsic curvature and not general derivatives thereof---prevents the outgoing extrinsic curvature from involving any of these new tensor structures.
Fluid variables end up only entering the scattering map through scalar invariants, and no independent tensorial structure survives in the classification.
It would be interesting to determine whether models with multiple matter fields admit scattering maps in which $K^+$ has tensorial structures that violate the scaling of the densitized anisotropy $|g^{\pm}|^{1/2}\Kcirc^{\pm}$.
}


\paragraph{Further generalization.} 

The construction also applies if the scattering map is only defined in a subset of the possible values. 
In particular, if we allow the scattering map to undefined at the isotropic point $\Kcirc^-=0$ (or merely non-differentiable), then an additional freedom appears, in the form of a scalar function~$\psi$ of the invariants~$\tau_m^-$, only, satisfying the \emph{scale-invariance} property
\bel{scale-invariance-psi}
\sum_{m=2}^d m \tau_m^- \frac{\del\psi}{\del\tau_m^-} = \psi .
\ee
The identities in~\eqref{anisotropic-map-fluid} are unchanged, except that the metric undergoes an anisotropic rescaling, expressed as a matrix exponential of traceless parts of the powers $(\Kcirc^-)^{m-1}$: 
\be
g^+ = \omega^{2/d} \exp\biggl( \sum_{m=2}^d 2 m \frac{\partial\psi}{\partial\tau_m^-} \bigl( (\Kcirc^-)^{m-1} \bigr)^\circ \biggr) g^- .
\ee
It would be interesting to see whether a similar relaxation occurs in the Einstein--scalar field case, with the appearance of a purely geometric parameter analogous to~$\psi$ that does not affect the matter sector.
The viability of such an extension should be assessed by considering specific small-scale modeling of the bounces. 


\section{Toward a full classification of interface conditions}
\label{section-6}

{\rouge

\paragraph{Interfaces and junction conditions.}

The present paper addresses the non-uniqueness arising in macroscopic
models of self-gravitating fluids with discontinuities or blow-up
singularities. Our starting point is the theory of sharp interfaces,
where undercompressive shocks and phase boundaries require, beyond the
classical conservation laws, an additional \emph{kinetic relation}
selecting the physically relevant solution. We extend this viewpoint
to the relativistic and self-gravitating regime, and to geometric singularities.

Causality arguments based on the acoustic and light cones distinguish
two classes of interfaces: \emph{compressive} and
\emph{undercompressive}. In the compressive case, when the geometry
remains regular and no phase transition occurs, the classical
junction conditions suffice: one combines the Israel conditions on the
geometry~\cite{Israel} (see also~\cite{MarsSenovilla} together with the mathematical presentation in \cite{LeFloch-Mardare})
with the relativistic fluid jump conditions of
Lichnerowicz~\cite{Lichne}. By contrast, geometric singularities and
undercompressive fluid interfaces require additional relations.

Our proposal is to encode these extra conditions in a
\emph{scattering law}. This notion combines ideas from cosmological
singularity theory~\cite{LLV-1a,LLV-1b,LLV-2} and phase transition
dynamics~\cite{AK1,AK2,LeFloch-book,LeFloch-Oslo,Truskinovsky}, while separating universal
constraints---covariance, locality, and causality---from
model-dependent relations reflecting the microscopic physics of the
interface. It thus provides a unified basis for classifying admissible
junction conditions for geometric singularities, relativistic shocks,
and phase transitions.

\paragraph{Microphysical input and concrete models.}

The scattering maps introduced here encode, at an effective level,
physical processes below the resolution of the Einstein--Euler
description. They supply the additional constitutive input needed
where the field equations alone do not determine a unique
continuation. Although their precise form depends on the underlying
microphysics, admissibility is already strongly constrained by
covariance, causality, and ultra-locality.

A concrete model is then obtained by specifying the physical setting,
the interface hypersurface, and the matter model, typically through an
equation of state, and by selecting a specific scattering law within the
admissible class. In cosmological applications, this prescribes how
geometric and fluid variables are matched across a bounce; in
astrophysical applications, it describes transmission across an
interface separating distinct phases of matter.  It would be interesting
to work out scattering maps in both settings, similar to the
map~\eqref{preBigBang-map} for the pre-Big-Bang scenario (which concerns
the Einstein--scalar system), and to how kinetic relations in
non-relativistic fluids can be determined from models with viscosity and
surface tension (or capillarity)~\cite{BedjaouiLF,FanSlemrod,Shearer1986}.
Our classification
results are intended as a filter for the resulting scattering maps.

}

{\rouge

\paragraph{On numerical relativity and stability.}

The present paper is analytical in nature and does not address
numerical implementation. Nevertheless, the scattering-law framework
is naturally compatible with standard methods in numerical relativity
and computational fluid dynamics.
It should enter the local
resolution of interface problems involving shocks, phase boundaries, or
singular hypersurfaces, for instance, for Riemann solvers in
finite-volume methods~\cite{LeFlochThanh,LeFlochMohamadian}.
A numerical implementation could rely on techniques for tracking interfaces.

We do not undertake a perturbative stability analysis either. (See, for instance, \cite{LiuZumbrun} for a study of the stability of undercompressive shocks.) Stability is constrained by the hyperbolic structure of the Einstein--Euler system: causality and ultra-locality ensure that
transmission across the interface is compatible with the relevant
characteristic directions and does not introduce spurious propagation
effects.
We conjecture that, aside from certain metastable phase transitions, an interface satisfying our scattering laws will remain stable under small perturbations.
A detailed study, especially in bounce scenarios, lies beyond the scope of the present work.

\paragraph{Concluding remarks.}

The scattering-law methodology developed here suggests several
directions for further research. A first priority is to investigate realistic equations of state, especially in astrophysical regimes, and to
determine how additional effects such as viscosity or
capillarity lead to concrete scattering maps.
In particular, non-convex equations of state in
relativistic hydrodynamics have been studied in recent
years~\cite{Berbel-Serna,Berbel-Serna-Marquina,Ibanez,Marquina-Serna-Ibanez,Rivieccio-Guerra-Ruiz-Font},
with possible observable consequences, for instance in
gravitational-wave signals from compact-binary mergers.
More broadly, sharp-interface models are especially appealing when widely disparate scales interact in the same dynamics, as in phase transitions in neutron-star matter
\cite{AnderssonComer,Blaschke,Christian,Ibanez}, in the formation of
trapped surfaces driven by gravitational or shock waves
\cite{Alvarez-Gaume}, or in the dynamics of galactic spiral arms
\cite{Kim}. It would also be of interest to extend the present
analysis to settings involving modified or quantum gravity
\cite{Ashtekar:2009a,De-Cesare,KOST,SteinhardtTurok2004,Wilson-Ewing}.

}


\ack{Both authors gratefully acknowledge financial support from the Simons Center for Geometry and Physics, Stony Brook University. 
This research was initiated when the first author (BLF) was a research fellow at the Princeton Center for Theoretical Science (PCTS), Princeton University, and the second author (PLF) was a visiting research fellow at the Courant Institute of Mathematical Sciences, New York University. During the completion of this paper, the authors were both supported by Projet-ANR-23-CE40-0010 \emph{``Einstein constraints: past, present, and future''} funded by the Agence Nationale de la Recherche (ANR),  
and by HORIZON-MSCA-2022-SE/101131233 funded by the European Research Council (ERC).}



\appendix 

\renewcommand \theequation {\Alph{section}\hskip2pt\arabic{equation}}

\section{Hugoniot curves for relativistic fluid boundaries}
\label{appendix-A}

\paragraph{Parametrization of the velocity field and interface speed.}

We are interested in the local properties at a point $p \in \Mcal$ of a single timelike jump discontinuity, which we shall describe in an orthonormal frame $\del_j$, $j=0,\dots,d$ at this point.  At the relevant order, orthonormality is equivalent to working with the Minkowski metric $g = - (dt)^2 + \sum_j (dx^j)^2$, which simplifies the presentation in this section since we can then easily raise or lower indices.
We denote by~$V$ the spacelike unit normal one-form to the discontinuity hypersurface.
To ease comparison with the well-known case of non-relativistic fluids in one spatial dimension, we rotate the spatial frame vectors to arrange $V = (V_0, V_1, 0, \ldots, 0)$, while keeping the vector $\del_0=\del_t$ unchanged as is necessary for possible applications to evolution problems.
Thanks to the normalization $- (V_0)^2+ (V_1)^2 = 1$, the normal~$V$ is equivalently described by the \emph{interface rapidity} $Z\in(-\infty,+\infty)$, with
\be
Z = \frac{1}{2} \log\frac{-V_0+V_1}{V_0+V_1} , \qquad
V_0 = - \sinh Z , \qquad
V_1 = \cosh Z .
\ee
By convention, after fixing an orientation, we can assume that 
the interface is \emph{moving toward} the  region $\phaseI$ if $z<0$, 
or \emph{moving toward} the region $\phaseII$ if $z>0$, while it is \emph{stationary} if $z=0$.

The fluid velocity (on both sides of the interface) decomposes as $(u^\alpha) = (u^\perp, u^\parallel)$ into perpendicular and parallel components $u^\perp := (u^0, u^1)$ and $u^\parallel := (u^a)$ ($a=2, \ldots d$) that lie along $\Span\{\del_t,V\}$ and along a time slice of the interface hypersurface, respectively.
Given the normalization $-(u^0)^2 + (u^1)^2 + |u^\parallel|^2 = -1$, it is natural to define the (scalar) \emph{perpendicular fluid rapidity} $W\in(-\infty,+\infty)$, with
\be
W = \frac{1}{2} \log\frac{u^0+u^1}{u^0-u^1} , \qquad
u^0 = (1 + |u^\parallel|^2)^{1/2} \cosh W , \qquad
u^1 = (1 + |u^\parallel|^2)^{1/2} \sinh W ,
\ee
so that the (unit, by definition) velocity vector $u$ is parametrized in (a one-to-one manner) by the scalar $W$ and the spacelike vector $u^\parallel$ orthogonal to~$\{\del_t,V\}$.

The rapidities are evaluated with respect to a choice of reference timelike vector~$\del_t$.
In fact, the only relativistic invariant combination of the unit vectors $u,V$ is the \emph{(scalar) perpendicular velocity} $v \in (-\infty, +\infty)$ defined by
\bel{equa-define-v}
v := V_\alpha u^\alpha
= V \cdot u^\perp
= V_0 u^0 + V_1 u^1
= (1+|u^\parallel|^2)^{1/2} \sinh(W-Z) .
\ee
In the non-relativistic limit (small $W,Z,u^\parallel$), the rapidities $Z$ and~$W$ reduce to the speeds of the interface and fluid (perpendicular to the interface), respectively, and $v$ reduces to their difference, which is a Galilean invariant.


\paragraph{Decomposition of the Rankine--Hugoniot relations.}

We split the jump relations into three parts: along $V$, along the (future-oriented timelike) vector~$\tV$ with components $(\tV^\alpha)=(V^1,V^0,0,\dots)$, which is tangent to the interface, and along the other tangent directions $\del_j$, $j=2,\dots,d$.
Contracting~\eqref{equa-RHII} with~$V^\beta$ gives us
\be
\aligned
& \bigl(\mu_\phaseII + p_\phaseII(\mu_\phaseII) \bigr)  (u_\phaseII\cdot V)^2
- \bigl(\mu_\phaseI + p_\phaseI(\mu_\phaseI) \bigr) (u_\phaseI\cdot V)^2
+ \bigl(p_\phaseII(\mu_\phaseII) - p_\phaseI(\mu_\phaseI) \bigr)  = 0,
\endaligned
\ee
which can be written in terms of rapidities and of $\pi^\parallel_\phase =(\mu_\phase + p_\phase(\mu_\phase))(1+|u_\phase^\parallel|^2) > 0$ for $\phase=\phaseI,\phaseII$.

\begin{claim}[Rankine--Hugoniot (1)]
\label{claim1} 
With $p_\phaseII = p_\phaseII(\mu_\phaseII)$ and $p_\phaseI = p_\phaseI(\mu_\phaseI)$, one has 
\[
\sinh^2(W_\phaseII-Z)
= (\pi_\phaseI^\parallel/\pi_\phaseII^\parallel) \sinh^2(W_\phaseI-Z)
+ (p_\phaseI - p_\phaseII) / \pi_\phaseII^\parallel .
\]
\end{claim}

Next, we contract~\eqref{equa-RHII} with the (future-oriented timelike) vector~$\tV$ with components $(\tV^\alpha)=(V^1,V^0,0,\dots)$, which is tangent to the interface.
We find that $(\mu+p)(u\cdot V)(u\cdot\tV)$ is the same on both sides of the interface.  Using the identity $u\cdot\tV = - (1+|u^\parallel|^2)^{1/2} \cosh (W-Z)$, and $2 \sinh(a) \cosh (a) = \sinh(2a)$, we obtain the following.

\begin{claim}[Rankine--Hugoniot (2)]
\label{claim2}
One has
$
\sinh(2(W_\phaseII-Z))
= (\pi_\phaseI^\parallel/\pi_\phaseII^\parallel) \sinh(2(W_\phaseI -Z)) .
$
\end{claim}


The last relation is \eqref{equa-RHII}~with $\beta$ replaced by $a=2, \ldots, d$, or alternatively the projection of~\eqref{equa-RHII} orthogonal to $\{\del_t,V\}$, which yields a relation for the parallel velocity, namely
\be
\aligned
u^\parallel_\phaseII  \, v_\phaseII  
=  u^\parallel_\phaseI  \, v_\phaseI \, 
\frac{\mu_\phaseI + p_\phaseI}{\mu_\phaseII+p_\phaseII}.
\endaligned
\ee

\begin{claim}[Rankine--Hugoniot (3)]
\label{claim3}
One has
\[
\sinh(W_\phaseII-Z) \, (1+|u_\phaseII^\parallel|^2)^{-1/2} \, u^\parallel_\phaseII
= (\pi_\phaseI^\parallel / \pi_\phaseII^\parallel) \, \sinh(W_\phaseI-Z) (1+|u_\phaseI^\parallel|^2)^{-1/2} u_\phaseI^\parallel.
\]
\end{claim}


\paragraph{A parametrization of Hugoniot curves for phase boundaries.}

On each side of the interface, our main variables are $\Omega = (\mu, W, u^\parallel)$, that is, two real functions and a $(d-1)$-vector, while the interface rapidity $Z$ is also one of the degrees of freedom; in total,  
we have derived $1+ (d-1) + 1= d+1$ jump relations. 
We regard  $\Omega_\phaseI = (\mu_\phaseI, W_\phaseI, u_\phaseI^\parallel)$ as a \emph{fixed state}, and we are now in a position
to introduce the  \underline{Hugoniot set} $\Hcal(\Omega_\phaseI)$
consisting, by definition, of all states $\Omega_\phaseII = (\mu_\phaseII, W_\phaseII, u_\phaseII^\parallel)$
satisfying the Rankine--Hugoniot relations in Claims 1, 2, and 3 \emph{for some interface rapidity}~$Z$.

Observe that if the fluid on one side has the same rapidity as the interface then $W_\phaseI = W_\phaseII = Z$ (by Claim~\ref{claim2}), making Claim~\ref{claim3} vacuuous, and Claim~\ref{claim1} then implies $p_\phaseI(\mu_\phaseI) = p_\phaseII(\mu_\phaseII)$.  This case describes a \emph{contact interface} across which the parallel velocity~$u^\parallel$ may jump arbitrarily, and the set $\Hcal(\Omega_\phaseI)$ contains a $(d-1)$-dimensional subspace.

Under non-degeneracy assumptions, it is known from the general theory of systems of nonlinear hyperbolic equations that $\Hcal(\Omega_\phaseI)$ consists of the union of $d+1$ \emph{curves} which emerge from a base point satisfying $(p_\phaseII(\mu_\phaseII),W_\phaseII,u_\phaseII^\parallel)=(p_\phaseI(\mu_\phaseI),W_\phaseI,u_\phaseI^\parallel)$.
The aforementioned subspace arises due to the parallel velocity components having equal wave speeds~\eqref{waves-multiplicity}.  The standard local analysis for the phase transition model under consideration carries over for the remaining variables, leading to two curves starting at the base point.  Since we are interested in possibly ``large'' jumps 
in order to encompass phase transitions, expliciting these curves is worthwhile.

We henceforth assume $W_\phaseI\neq Z$ and (equivalently) $W_\phaseII\neq Z$.
Dividing the identity in Claim~\ref{claim3} by Claim~\ref{claim2} yields
\bel{upar-A6}
u^\parallel_\phaseII
=
\frac{\cosh(W_\phaseII-Z) (1+|u_\phaseII^\parallel|^2)^{1/2}}{\cosh(W_\phaseI-Z) (1+|u_\phaseI^\parallel|^2)^{1/2}} \, u_\phaseI^\parallel .
\ee
The following identities can be checked by rewriting $p_\phaseI - p_\phaseII$ using Claim~\ref{claim1}, then $\pi_\phaseI^\parallel/\pi_\phaseII^\parallel$ using Claim~\ref{claim2}:
\bel{pi12-A7}
\aligned
1 + (p_\phaseI - p_\phaseII) / \pi_\phaseII^\parallel 
& = \cosh(W_\phaseII-W_\phaseI) \cosh(W_\phaseII-Z) / \cosh(W_\phaseI-Z) ,
\\
1 - (p_\phaseI - p_\phaseII) / \pi_\phaseI^\parallel 
& = \cosh(W_\phaseII-W_\phaseI) \cosh(W_\phaseI-Z) / \cosh(W_\phaseII-Z) .
\endaligned
\ee
The norm-squared of~\eqref{upar-A6} and ratio of~\eqref{pi12-A7} involve the same hyperbolic cosines, hence
\[
\bigl(1 - (p_\phaseI - p_\phaseII) / \pi_\phaseI^\parallel\bigr) (1+|u_\phaseI^\parallel|^2)|u^\parallel_\phaseII|^2
= \bigl( 1 + (p_\phaseI - p_\phaseII) / \pi_\phaseII^\parallel \bigr) (1+|u_\phaseII^\parallel|^2)|u_\phaseI^\parallel|^2 .
\]
Combined with the definitions $\pi_\phase^\parallel = (\mu_\phase+p_\phase)(1+|u^\parallel_\phase|^2)$, this eventually simplifies to a relation between norms of~$u^\parallel_\phase$, hence between the vectors themselves (due to~\eqref{upar-A6}),
\bel{Hugoniot-upar}
u^\parallel_\phaseII = \Bigl(\frac{(\mu_\phaseI+p_\phaseI)(\mu_\phaseII+p_\phaseI)}{(\mu_\phaseI+p_\phaseII)(\mu_\phaseII+p_\phaseII)}\Bigr)^{1/2} u_\phaseI^\parallel .
\ee
Lastly, taking the product of the equations in~\eqref{pi12-A7} eliminates~$Z$ and allows us to determine the rapidity~$W_\phaseII$, except for a sign of $W_\phaseII-W_\phaseI$,
\bel{Hugoniot-W}
\sinh^2(W_\phaseII-W_\phaseI)
= \frac{(p_\phaseI - p_\phaseII) \bigl(\mu_\phaseI + (\mu_\phaseI + p_\phaseI) |u^\parallel_\phaseI|^2 - \mu_\phaseII - (\mu_\phaseII + p_\phaseII) |u^\parallel_\phaseII|^2\bigr)}{(\mu_\phaseI + p_\phaseI) (\mu_\phaseII + p_\phaseII)(1 + |u^\parallel_\phaseI|^2) (1 + |u^\parallel_\phaseII|^2)}
\ee
in which $u^\parallel_\phaseII$ is a short-hand for~\eqref{Hugoniot-upar}.  Under the monotonicity condition $\sgn(p_\phaseI(\mu)-p_\phaseII(\tmu)) = \sgn(\mu-\tmu)$, these two equations describe two Hugoniot curves (for $W_\phaseII-W_\phaseI\gtrless 0$) parametrized by~$\mu_\phaseII$ that manifestly connect at the base point $(p_\phaseII,W_\phaseII,u^\parallel_\phaseII)=(p_\phaseI,W_\phaseI,u^\parallel_\phaseI)$.  They simplify drastically if the chosen frame has $u^\parallel_\phase=0$, and they admit a good non-relativistic limit $p_\phase/\mu_\phase, W_\phase,|u^\parallel_\phase|\ll 1$.

For given states $\Omega_\phaseI$, $\Omega_\phaseII$ satisfying these relations, the Rankine--Hugoniot relations hold provided the interface rapidity~$Z$ obeys $\sgn(Z-W_\phaseI) = \sgn(\mu_\phaseII-\mu_\phaseI) \sgn(W_\phaseII-W_\phaseI)$ and
\bel{interface-rapidity-value}
(\mu_\phaseI-p_\phaseI - \mu_\phaseII+p_\phaseII) \sinh^2(Z-W_\phaseI)
= \frac{(p_\phaseI-p_\phaseII)(\mu_\phaseII+p_\phaseI)}{(\mu_\phaseI+p_\phaseI)(1+|u^\parallel_\phaseI|^2)} .
\ee
An interesting case is that of stiff fluids~\eqref{hyperbolic-eos-1} where the sound speed in each phase coincides with the light speed, namely $p_\phase(\mu)=q_\phase^\star+\mu$ for some constitutive constants~$q_\phase^\star$ for $\phase=\phaseI,\phaseII$.
Observe that $\mu_\phaseI-p_\phaseI - \mu_\phaseII+p_\phaseII = q_\phaseII^\star-q_\phaseI^\star$.
If this constant is non-zero, one finds solutions of~\eqref{interface-rapidity-value} provided $p_\phaseI-p_\phaseII$ has the same sign as $q_\phaseII^\star-q_\phaseI^\star$ in the range of densitites of interest.  Such interfaces describe phase boundaries separating stiff fluids with distinct constitutive constants, which propagate at subsonic speeds, namely are timelike.
In contrast, for $q_\phaseI^\star=q_\phaseII^\star$, the two pressure laws coincide and the left-hand side degenerates, so that $p_\phaseII=p_\phaseI$ hence $W_\phaseII=W_\phaseI$ by~\eqref{Hugoniot-W}.
The Hugoniot curves are thus ruled out in this case, which reflects the fact that a single stiff fluid admits no shock, but only contact interfaces propagating at light speed.

Incidentally, if $\mu_\phaseI-p_\phaseI\neq\mu_\phaseII-p_\phaseII$, \eqref{Hugoniot-W}--\eqref{interface-rapidity-value}
can be reformulated invariantly as ($\phase=\phaseI,\phaseII$)
\bel{dot-by-RH}
(u_\phaseI \cdot u_\phaseII)^2
= \frac{(\mu_\phaseI+p_\phaseII)(\mu_\phaseII+p_\phaseI)}{(\mu_\phaseI+p_\phaseI)(\mu_\phaseII+p_\phaseII)} ,
\qquad
(V\cdot u_\phase)^2
= \frac{(p_\phaseI-p_\phaseII)(\mu_\phaseI+\mu_\phaseII-\mu_\phase+p_\phase)}{(\mu_\phaseI-p_\phaseI - \mu_\phaseII+p_\phaseII)(\mu_\phase+p_\phase)} .
\ee


\section{Irrotational stiff fluids}
\label{appendix-B}

\paragraph{Euler equations.}

To relate our investigation of scattering maps for the Einstein-scalar equations summarized in \autoref{section-2} with the Einstein--Euler case studied in the present paper, we explain now how the fluid reduces in a special case to a scalar field.
Consider a stiff fluid, whose pressure, by definition, is equal to its mass-energy density, namely $p=\mu$, so that the sound speed coincides with the light speed normalized to~$1$.
The stress-energy tensor~\eqref{eq:45} is then $T_{\alpha\beta} =  2 \mu \, u_\alpha u_\beta + \mu \, g_{\alpha\beta}$,
where $u$ is the future-oriented vector satisfying $u^\alpha u_\alpha = -1$.

The twice-contracted Bianchi identities imply the Euler equations  
$\nabla^\alpha T_{\alpha\beta} = 0$, which read
\bel{eqB1}
(u^\alpha \, \nabla_\alpha \mu) \, u^\beta + \mu \, (\nabla_\alpha u^\alpha) \, u^\beta  
+ \mu \, u^\alpha \, \nabla_\alpha u^\beta + \frac{1}{2} \nabla^\beta \mu = 0. 
\ee 
Contracting~\eqref{eqB1} with $-u_\beta$ and using $u_\beta \nabla_\alpha u^\beta = 0$, we obtain
$
\frac{1}{2} u^\alpha \nabla_\alpha \mu + \mu \, \nabla_\alpha u^\alpha = 0. 
$
Consequently, assuming that the density is bounded away from zero  and setting 
$\nu : = \frac{1}{2}  \log \mu$ we arrive at an evolution equation for~$\nu$, namely
$u^\alpha \nabla_\alpha \nu + \nabla_\alpha  u^\alpha = 0$. 
On the other hand, projecting~\eqref{eqB1} orthogonally to~$u$ by the projection operator $H_{\alpha\beta} := g_{\alpha\beta} + u_\alpha u_\beta$ we find the evolution equation associated with the velocity: 
$\mu \, u^\alpha  \nabla_\alpha u^\beta + \frac{1}{2} H^{\alpha\beta} \nabla_\alpha \mu = 0$.
Altogether,
\be\label{unablau}
u^\alpha \nabla_\alpha \nu + \nabla_\alpha  u^\alpha = 0, \qquad
u^\alpha \nabla_\alpha u^\beta + H^{\alpha\beta} \nabla_\alpha \nu = 0.
\ee


\paragraph{Irrotational stiff fluid as a scalar field.}
Consider now the irrotational case, where the fluid velocity is the (normalized) gradient of a scalar potential~$\psi$, 
\bel{equa-A1} 
u^\alpha = e^{-f} \nabla^\alpha\psi , \qquad e^{-f} := (-\nabla_\beta \psi \, \nabla^\beta \psi)^{-1/2} .
\ee
By contracting $\nabla_\alpha u_\beta-\nabla_\beta u_\alpha = u_\alpha \nabla_\beta f - u_\beta \nabla_\alpha f$ with $u^\alpha$ and using that $u^\alpha\nabla_\beta u_\alpha=0$, the velocity evolution equation~\eqref{unablau} can be rewritten as
$H^{\alpha\beta} \nabla_\alpha \nu = \nabla_\beta f + u^\alpha \nabla_\alpha f u^\beta = H^{\alpha\beta} \nabla_\alpha f$,
from which we deduce that $\nabla (\nu - f)$ is parallel to~$u$.
The vector field $e^\nu u=\mu^{1/2}u$ thus has vanishing curl,
\be
\nabla_\alpha(e^\nu u_\beta) - \nabla_\beta(e^\nu u_\alpha)
= e^\nu (u_\beta \nabla_\alpha - u_\alpha \nabla_\beta)(\nu - f) = 0 ,
\ee
hence it is the gradient of a scalar field~$\phi$, namely (with a convenient normalization and sign)
\be\label{2mu12ualpha}
(2\mu)^{1/2} u_\alpha = \pm \nabla_\alpha\phi .
\ee

The fluid's stress-energy tensor is then $T_{\alpha\beta} =  \nabla_\alpha\phi \nabla_\beta\phi - \frac{1}{2} (\nabla_\gamma\phi\nabla^\gamma\phi) \, g_{\alpha\beta}$, namely that of a massless scalar field, which is minimally coupled to the Einstein equations for the metric, that is, $R_{\alpha\beta} - 8\pi\, \nabla_\alpha\phi\nabla_{\beta} \phi = 0$.
The Euler equations for the fluid coincide with the scalar wave equation since both arise as the conservation equation $\nabla^\alpha T_{\alpha\beta} = 0$.
In detail, the fluid evolution equations in~\eqref{unablau} are the wave equation
$2^{1/2} \nabla_\alpha  (e^\nu u^\alpha) = \nabla_\alpha\nabla^\alpha\phi = 0$ and the compatibility equation $\nabla_\alpha\nabla_\beta\phi=\nabla_\beta\nabla_\alpha\phi$.

Conversely, a massless scalar field can be seen as an irrotational stiff fluid, provided its gradient~$\nabla\phi$ is timelike (with the appropriate sign in~\eqref{2mu12ualpha}) so that fluid variables $\mu=-\frac{1}{2}\nabla_\alpha\phi\nabla^\alpha\phi>0$ and $u_\alpha$ are well-defined.


\paragraph{From Einstein-scalar scattering maps to fluids.}

In the quiescent regime near spacelike gravitational singularities (\autoref{section-2}), the time derivative $\del_t\phi$ dominates spatial derivatives hence $\nabla\phi$ is timelike.  In terms of fluid variables, the asymptotic profiles~\eqref{gKphi-star} yield (with $a=1,\dots,d$)
\be
\mu_*(t) = \frac{1}{2t^2} \bigl((\phi_0^{\pm})^2 + o(1)\bigr) , \qquad
u_{*a}(t) = t(\phi_0^{\pm})^{-1} \Bigl(\del_a \phi_0^{\pm}\log|t/t_*| + \del_a \phi_1^{\pm}\Bigr) + o(t) ,
\ee
and $u_{*0}(t) = 1 + o(1)$.
Up to lower-order terms, this matches the asymptotic profiles~\eqref{gKmuu-star},
with $\mu^{\pm}= (\phi_0^{\pm})^2/2$ and $u^{\pm} = (\phi_0^{\pm})^{-1} \del_a \phi_1^{\pm}$.

Since the sets of singular data $(g^{\pm},K^{\pm},\mu^{\pm},u^{\pm})$ do not include~$\phi_1^{\pm}$ but only its spatial derivatives,
a scattering map $\Scal:\ (g^-,K^-,\phi_0^-,\phi_1^-)\longmapsto (g^+,K^+,\phi_0^+,\phi_1^+)$ fails to be local in terms of the fluid variables, unless shifting $\phi_1^-$ by a constant simply shifts $\phi_1^+$ by a constant.  We classified such shift-covariant maps in \cite[Theorem 5.4]{LLV-1a} in $d=3$ dimensions.
\begin{itemize}
\item \textbf{Isotropic scattering maps} are~\eqref{Siso-expr} with coefficients $\Delta_n$ independent of $\phi_1^-$, hence expressible in terms of scalar invariants of~$K^-$ (and the sign of~$\phi_0^-$).  In fluid variables, this states $(K^+,\mu^+,u^+)=\bigl((1/d)\delta,(d-1)/(2d),0\bigr)$, with a suitable~$g^+$.
\item \textbf{Anisotropic scattering maps} are given by~\eqref{Sani-expr} with $\phi_0^+/r(\phi_0^+) = \eps \zeta^{-1}\phi_0^-/r(\phi_0^-)$ and $\phi_1^+=\zeta f(\chi_m,\phi_0^-)+\zeta\phi_1^-$ for some sign $\eps$, real parameter $\zeta\neq 0$, and function~$f$ of the scalar invariants.
This leads to $1/\mu^+ = \zeta^2/\mu^- + 2 (1-\zeta^2) d/(d-1)$ and
$u^+_a = \zeta (\phi_0^+)^{-1} (\del_af(\chi_m,\phi_0^-) + \phi_0^- u^-_a)$.
\end{itemize}
For the latter maps, selecting $f$~to be constant ensures that $u^+$~depends ultra-locally on $(g^-,K^-,\mu^-,u^-)$.
In that case, $g^+=c^2(\mu^-/\mu^+)^{1/d}g^-$ for some constant parameter $c>0$, and the maps reproduce the anisotropic maps of \autoref{thm:fluid-scattering-map} with $\gamma = \eps c^d |\zeta|$ and a scale factor
\be
\omega = c^d \sqrt{\zeta^2 + \tfrac{2d}{d-1} (1-\zeta^2) \mu^-}.
\ee
Shift-covariant scattering maps with arbitrary functions~$f$ yield a fluid speed~$u^+$ that depends \emph{locally but not ultra-locally} on the incoming data.
An interesting relaxation of the notion of ultra-locality for fluids is thus to allow the fluid velocity~$u^+$ to depend on (first) derivatives of~$(K^-,\mu^-)$ in addition to $(g^-,u^-)$ without derivatives.

\end{document}